%% file: ejecta.tex
\documentclass{aa}
\usepackage{natbib}
\usepackage{graphicx}
\usepackage{txfonts}
\usepackage{psfig}

\newcommand{\ra}[3]{\mbox{R.A.(J2000)}={#1}$^{{\rm h}}${#2}$^{{\rm m}}${#3}$^{{\rm s}}$}
\newcommand{\dec}[3]{\mbox{Dec(J2000)}={#1}\degr{#2}\arcmin{#3}\arcsec}

\newcommand{\xmm}{\emph{XMM-Newton}}
\newcommand{\chandra}{\emph{Chandra}}
\newcommand{\chisq}{$\chi^2$}
\newcommand{\cm}[1]{~cm$^{#1}$}
\newcommand{\cms}{~cm$^{-3}$\,s}

\newcommand{\ctb}{~cts\,s$^{-1}$\,bin$^{-1}$}
\newcommand{\ctam}{~cts\,s$^{-1}$\,arcmin$^{-2}$}
\newcommand{\e}[1]{10$^{#1}$}
\newcommand{\ee}[1]{\,$\times$\,10$^{#1}$}

\newcommand{\hi}{H\,{\sc i}}

\newcommand{\msun}{M$_{\odot}$}
\newcommand{\nh}{N$_{\rm H}$}

\newcommand{\she}{S\,\textsc{xv}}
\newcommand{\sh}{S\,\textsc{xvi}}

\begin{document}

\title{XMM-Newton observations of the supernova remnant IC~443:}

\subtitle{II.~evidence of stellar ejecta in the inner regions. }

\author{Eleonora Troja\inst{1,2}
        Fabrizio Bocchino\inst{3}, 
	Marco Miceli\inst{3},
	\& Fabio Reale\inst{2,3}}

\institute{
INAF - Istituto di Astrofisica Spaziale e Fisica Cosmica, 
Sezione di Palermo, via Ugo la Malfa 153, 90146 Palermo, Italy
\and
Dipartimento di Scienze Fisiche ed Astronomiche,
Sezione di Astronomia, Universit\`a di Palermo, Piazza del
Parlamento 1, 90134 Palermo, Italy 
\and 
INAF-Osservatorio Astronomico di Palermo, Piazza
del Parlamento 1, 90134 Palermo, Italy}

\date{Received --, accepted --}

\titlerunning{Stellar ejecta in the SNR IC~443} 
\authorrunning{Eleonora Troja et al.}

\abstract
{}
{
We investigate the spatial distribution of the physical and chemical properties of the 
hot X-ray emitting plasma of the supernova remnant IC~443, 
in order to get important constraints on its ionization stage, 
on the progenitor supernova explosion, on the age of the remnant, 
and its physical association with a close pulsar wind nebula.
}
{We present XMM-Newton images of IC~443, median photon energy map, 
silicon and sulfur equivalent width maps, 
and a spatially resolved spectral analysis of a set of homogeneous regions.}
{The hard X-ray thermal emission (1.4--5.0~keV) of IC~443 displays a centrally-peaked morphology,
its brightness peaks being associated with hot (kT$>$1~keV) X-ray emitting plasma. 
A ring-shaped structure, characterized by high values of equivalent widths and
median photon energy, encloses the PWN. Its hard X-ray emission is
spectrally characterized by a collisional ionization equilibrium model, 
and strong emission lines of Mg, Si, and S,  requiring oversolar metal abundances. 
Dynamically, the location of the ejecta ring suggests an SNR age of $\sim$4,000 yr.
 The presence of overionized plasma in the inner regions of IC~443, 
addressed in previous works, is much less evident in our observations. }
{}


\maketitle

\section{Introduction}\label{sec:intro}

 The Galactic SNR IC~443 (G189.1+3.0) has an optical and radio 
double shell morphology \citep{leahy04,reich03,braun86,duin75}.
Signatures of interaction with a dense and 
complex environment are visible at radio \citep[e.g.][]{snell05,denoyer77}
and IR wavelengths \citep[and references therein]{neufeld08,rho01}.
In the northeast the shock front has been decelerated by the encounter
with a \hi\ cloud \citep{rho01,dickel89,denoyer78}.
A giant molecular cloud, mapped by \citet{cornett77},
is located in the foreground, and is interacting with the remnant
at several positions along the southern rim \citep[as schematically represented in Fig.~9 of][]{troja06}. 
IC~443 is also a source of $\gamma$-rays \citep{esposito96,sturner95}.
Interestingly, in the same region of sky where OH
(1720~MHz) maser emission is detected \citep{hewitt06,hoffman03,claussen97},
\citet{albert07} discovered a source of very high energy $\gamma$--ray emission,
possibly originated by the interactions between cosmic rays accelerated in IC 443 
and the dense molecular cloud.

Very interesting features and a peculiar morphology
are displayed in the X-ray band 
\citep{bykov08,bykov05,bocchino03b,asaoka94,wang92,petre88}.
The bulk of its X-ray emission is thermal, fairly described
with a two components ionization equilibrium model of temperatures
$\sim$0.2--0.3~keV (hereafter ``cold component'') and $\gtrsim$1.0~keV
(hereafter ``hot component''; \citealt{troja06,asaoka94,petre88}).
A source of very hard X-ray emission, identified as 
a pulsar wind nebula (PWN), lies close to the SNR's southern rim and
is possibly associated with the remnant \citep{gaensler06,bocchino01,olbert01}.

On the basis of its X-ray properties, as observed by \emph{ROSAT}, 
IC~443 has been classified as a mixed morphology (MM) SNR.
MM SNR main features were codified by \citet{mixed98} as follows: 
1)~the X-ray emission is thermal and centrally
peaked, with little or no evidence of a limb-brightened X-ray shell;
2)~it arises primarily from swept-up interstellar material, not from
ejecta; 3)~a~flat temperature radial profile, and a constant or
increasing density toward the remnant center; 4) a physical
association with dense molecular clouds.

\xmm\  high resolution observations of IC~443, 
presented by \citet{troja06}, resolved a limb-brightened shell structure 
in the very soft X-ray band (0.3--0.5~keV), 
and reported a first evidence of Mg, Si and S enhanced abundances.
In addition, \xmm\ and \chandra\ observations of several other MM SNRs
unveiled very steep X-ray brightness profiles and multi-component 
X-ray-emitting plasma with enhanced abundances \citep{lazendic06,shelton04a}.
More recent results suggest indeed a non negligible presence of metal 
enrichment gradients and stellar ejecta inside at least~$\sim$50\% 
of all known MM SNRs.

Strong Si and S lines in the IC~443 X-ray spectrum were 
first reported by \citet{petre88},
and then studied in detail by \citet{kawasaki02}. 
In the latter work the X-ray emission from IC~443 was described
by a two temperatures model, with
a 0.2~keV component in ionization non-equilibrium (NEI) 
and a 1.0~keV component in collisional ionization equilibrium (CIE).
Narrow Gaussians were used to reproduce the most prominent lines.
The elements abundances were all below the solar value, 
as expected for X-ray emission from the shocked ISM.
From the relative strengths of Si and S lines, \citet{kawasaki02}
derived an ionization temperature of $\sim$1.5~keV, significantly 
above the 1.0~keV continuum temperature. 
Those results led to the hypothesis of an overionized thermal plasma.

Using \xmm\ observations of the SNR IC~443, we have already presented 
a detailed analysis of the soft X-ray thermal emission, 
mainly originated from the shocked ISM (see \citealt{troja06}). 
Unlike \citet{kawasaki02} our spectral results gave a soft X-ray component 
(kT$\sim$0.2--0.3~keV) near the equilibrium condition ($\tau$$\sim$\e{12}\cms) 
in most of the analyzed regions.
The hot component (kT$\gtrsim$1.0~keV) was in full equilibrium and showed a 
high metallicity.
 
The present work focuses on the properties of the hot X-ray thermal component,
and is aimed at 1) 
addressing the high metal abundance plasma inside the
remnant and its ionization stage; 2) checking whether
the abundance pattern of the hot X-ray emitting plasma
is consistent with a core-collapse SN as progenitor,
as already hinted by the presence of the PWN.

The paper is organized as follows: 
observations are described in \S\ref{sec:data}; 
results from imaging and spectral analysis are presented in \S\ref{sec:results}.
In \S\ref{sec:discussion} we discuss their implications for
the remnant ionization stage (\S\ref{sec:plasma}), the SN progenitor (\S\ref{sec:sn}), 
and the age of IC~443 (\S\ref{sec:age}). 
A summary of our conclusions is presented in \S\ref{sec:end}.
 Throughout the paper, the quoted uncertainties
are at the 90\% confidence level \citep{lampton76}, unless otherwise stated.

\section{Observations and Data Reduction}\label{sec:data}

We analyzed four pointings \xmm\ observations of IC~443, 
performed during the Cal/PV phase on 2000 September 25-28 
and previously described in \citet{troja06} and \citet{bocchino03b}.
We added a more recent and very deep \xmm\ observation, taken on 2006 March 30
(ObsID~0301960101, P.I.~F.~Bocchino), and centered on the 
southeastern side of the remnant at \ra{06}{18}{04}, 
\dec{+22}{27}{33} \citep[see also][]{bykov08}.\\
Our work is focused on the analysis of EPIC MOS data.
We used the Science Analysis System software (SAS,version~7.0) 
for data processing and \textsc{xspec}~v.11.3 \citep{arnaud96} 
for spectral analysis.

Data were screened to remove spurious events and time intervals with heavy
proton flaring.
After the particle background removal,
the total exposures for the MOS cameras (MOS1+MOS2) 
were $\sim$170~ks and $\sim$140~ks
for the 2000 and 2006 data sets, respectively.

In order to estimate the Galactic background we used a set of blank sky 
observations performed as part of the Galactic Plane Survey
(GPS, P.I.~Parmar). We verified that, in the 
energy band (0.3--5.0~keV) selected for the analysis,
our results are not sensitive to the background subtraction 
(either GPS or high Galactic latitude blank fields).
We refer to our previous paper \citep{troja06} for more details about the
observations and the data reduction techniques we used.

\section{Results}\label{sec:results}
\subsection{X-ray Images}\label{sec:xima}


\begin{figure}
\centerline{\hbox{
     \psfig{figure=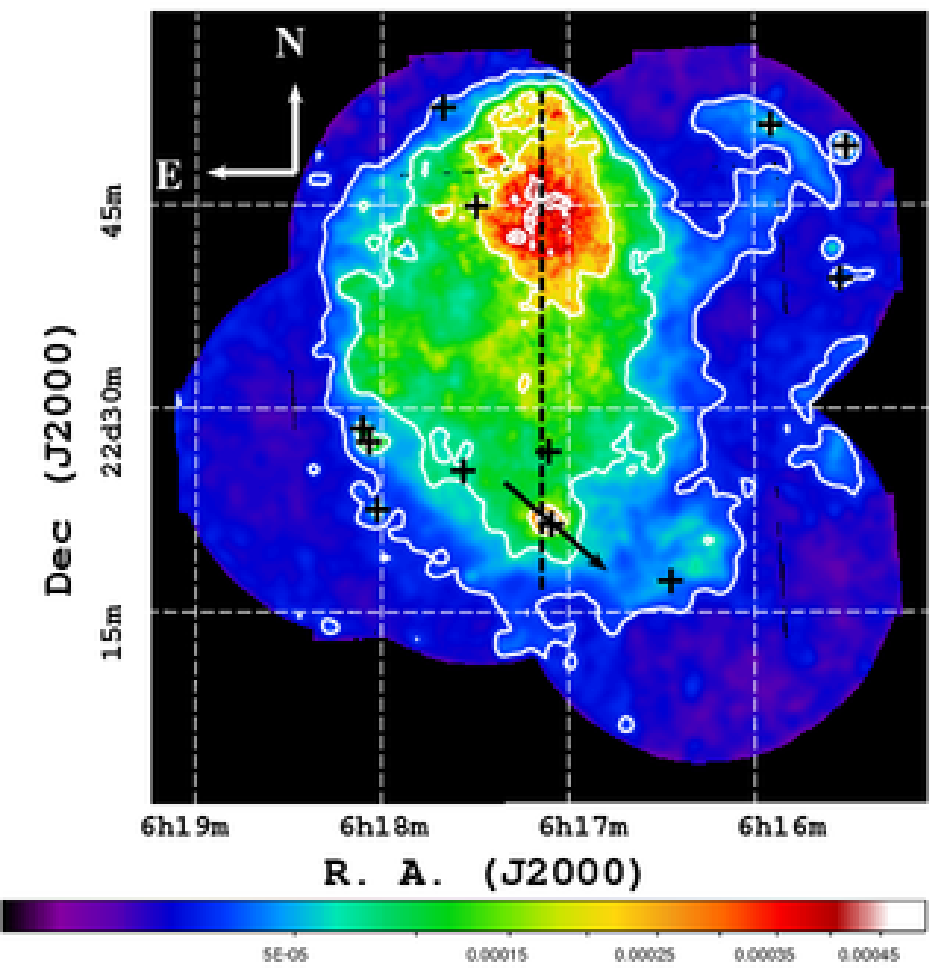,width=0.95\columnwidth}	 
     }}
\caption{X-ray count rate image of IC~443 in the 1.4--5.0 keV energy band.
       Units are\ctb.
       The image was adaptively smoothed using $\sigma_{min}$=5\arcsec\ 
       and $\sigma_{max}$=20\arcsec; the bin size is 5\arcsec.
       Contour levels correspond to 0.4, 1.0, 2.7, and 5.8$\times$10$^{-2}$\ctam.
       Black crosses mark the position of the 12 hard X-ray point sources
       identified by \citet{bocchino03b}.
       The dashed line traces the symmetry axis of the X-ray emission in the
       hard band; the black arrow indicates the putative direction of motion of
       the PWN, according to \citet{gaensler06}.}
\label{fig:xima}
\end{figure}

Fig.~\ref{fig:xima} shows the composite image of IC~443 in 
the hard 1.4-5.0 keV energy band, where the contribution of the cold
X-ray emitting plasma is negligible (see \citealt{troja06}). 
The image is background subtracted, exposure and vignetting corrected. 
We reported the position of the 12 hard X-ray point sources detected by
\citet{bocchino03b}. 

The hard band morphology does not resemble IC~443 emission in other bands
(optical, radio, IR and soft X-ray) and, on the plane of the image, it appears
inner concetrated with respect to the cold X-ray emitting plasma.

The hard X-ray thermal emission is dominated by an elongated
axis-symmetric structure, having its surface brightness peak in the
northeastern (NE) quadrant.  
The symmetry axis, traced by the dashed line in Fig.~\ref{fig:xima}, 
does not match with the plerion wind nebula (PWN) direction of motion
(-130$^o$ N through E, \citealt{gaensler06}), 
as indicated by the arrow in Fig.~\ref{fig:xima}.

It is worth noting that the X-ray surface brightness is higher 
($\Sigma$\,$\geq$\,0.012\ctam) in the subshell A
(following the naming convention of \citealt{braun86}; see also
Fig.~5 of \citealt{troja06}), 
which is associated with the presence of the neutral cloud in the NE,
while it appears strongly reduced ($\Sigma$\,$<$\,0.012\ctam)
in the western part (subshell B), where the blast wave is expanding 
in a more rarefied (n$_0$$\sim$0.25\,\cm{-3})
and homogeneous medium.

\subsection{Median Energy Map}\label{sec:median}

\xmm\ great sensitivity and collecting area enabled us to create a
high resolution image of the median photon energy $E_{50}$ in the
1.4--5.0~keV band. In the selected energy range the median energy is
mostly influenced by the parameters of the hot X-ray component, being
unaffected by spectral variations of the colder one. For a
plasma in ionization equilibrium and with solar abundances (model
MEKAL in {\sc xspec}) we verified that the $E_{50}$ distribution represents a
powerful tracer of temperature variations, especially for kT$\geq$1~keV.

\begin{figure}[h!]
\centerline{\hbox{
     \psfig{figure=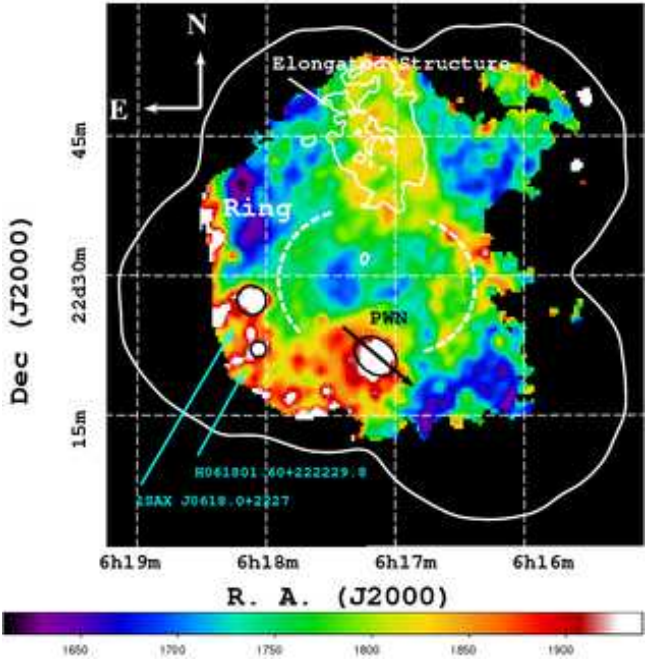,width=0.95\columnwidth}	 
     }}
\caption{
Median photon energy map in the hard band (1.4-5.0~keV). Units are eV.
Hard X-ray emission contours, corresponding to 
2.7 and 5.8$\times$10$^{-2}$\ctam, are overlaid. 
Noisy (S/N$<$3) regions inside the FOV, 
delimited by white solid line, were masked. 
The bin size is 20\arcsec and the
smoothing width is $\sigma$=20\arcsec.
 The two dashed semi-circles guide the eye
to notice a ring-shaped structure, which is more evident in the EW maps shown in Fig.~3.
The location of some sources, partially responsible for the high median energy in the SE region,
is indicated with green arrows.} 
\label{fig:q50}
\end{figure}

Fig.~\ref{fig:q50} shows the median energy map of the X-ray emitting
plasma in the 1.4-5.0 keV band. For a comparison, we overlaid
the brightest hard X-ray emission contours of Fig.~\ref{fig:xima}.
The white dashed line guides the eye to identify a ring-shaped
structure, which will be discussed in \S\ref{sec:ew}.

The median photon energy image clearly shows a north-to-south elongated structure,
spatially corresponding to the brightest emission features
in the hard energy band (contours in Fig.~\ref{fig:q50}). 
 The region of high median energy ($E_{50}$$\gtrsim$1.9~keV) along the southeast edge 
can be partially attributed to the emission of the PWN and of the other point sources \citep{bocchino03b},
but it appears to have a much larger extent, suggestive of an extended hard emission.
The nature of the hard X-ray source 1SAX J0618.0+2227 (also marked in Fig.~\ref{fig:q50}) 
has been extensively investigated  \citep{bykov08,bykov05,bocchino00} 
and interpreted in the framework of interaction of a molecular cloud
with a fast moving SN ejecta fragment. 
The study of the diffuse hard emission revealed in Fig.~\ref{fig:q50} 
is beyond the scope of this work, however we note that it 
may have a similar physical origin, probably arising from the interaction of the SNR 
with the structured cloud \citep{vandish93,burton90}.


\subsection{Equivalent Width Images}\label{sec:ew}

IC~443 X-ray emission is primarily thermal and
its spectrum is dominated by prominent atomic emission lines. 
Its global spectrum can be described with a phenomenological model,
consisting of two bremsstrahlung components for the continuum 
(kT$\sim$0.2 keV and kT$\sim$1.1 keV)
plus Gaussian features reproducing the line emission.

We explored the spatial distribution of the hard X-ray lines
emission 
by constructing equivalent width (EW) images,
often used as powerful tracers of stellar ejecta fragments in SNRs
\citep{hwang00,gotthelf01,park03a,cassam04,miceli06}. 

We selected an energy band for each emission line, 
combining the He$\alpha$ and Ly$\alpha$ emission of each element 
(e.g. Si\,XIII and Si\,XIV) in a single image to improve the statistic.
We chose the energy band that optimizes the signal to noise (S/N) ratio 
according to our best-fit phenomenological model \citep{cassam04}.

X-ray emission in the line energy band needs to be corrected for the
underlying continuum emission. The correction is made pixel by pixel
because the continuum may be spatially distributed in
a different way from the line emission. 
In order to account for the continuum contamination, 
we selected two continuum bands located on either side of the line energy band.
Their width was chosen to optimize the S/N ratio, according to the
recipe of \citet{cassam04}.

This method was successfully applied to silicon and sulfur lines.
We excluded the H-like Mg line at 1.47~keV since 
the underlying continuum emission cannot be correctly estimated. 
In this particular case, 
the low energy continuum band ($\sim$1.1~keV) is dominated by the soft spectral
component and still affected by the strong absorbing column gradients, 
due to the foreground molecular cloud.  

Line and continuum band passes, selected for each spectral line of interest,
are listed in Tab.~\ref{tab:bands}.

\input{tab1.tex}

\begin{figure}[!h]
\centerline{\hbox{
     \psfig{figure=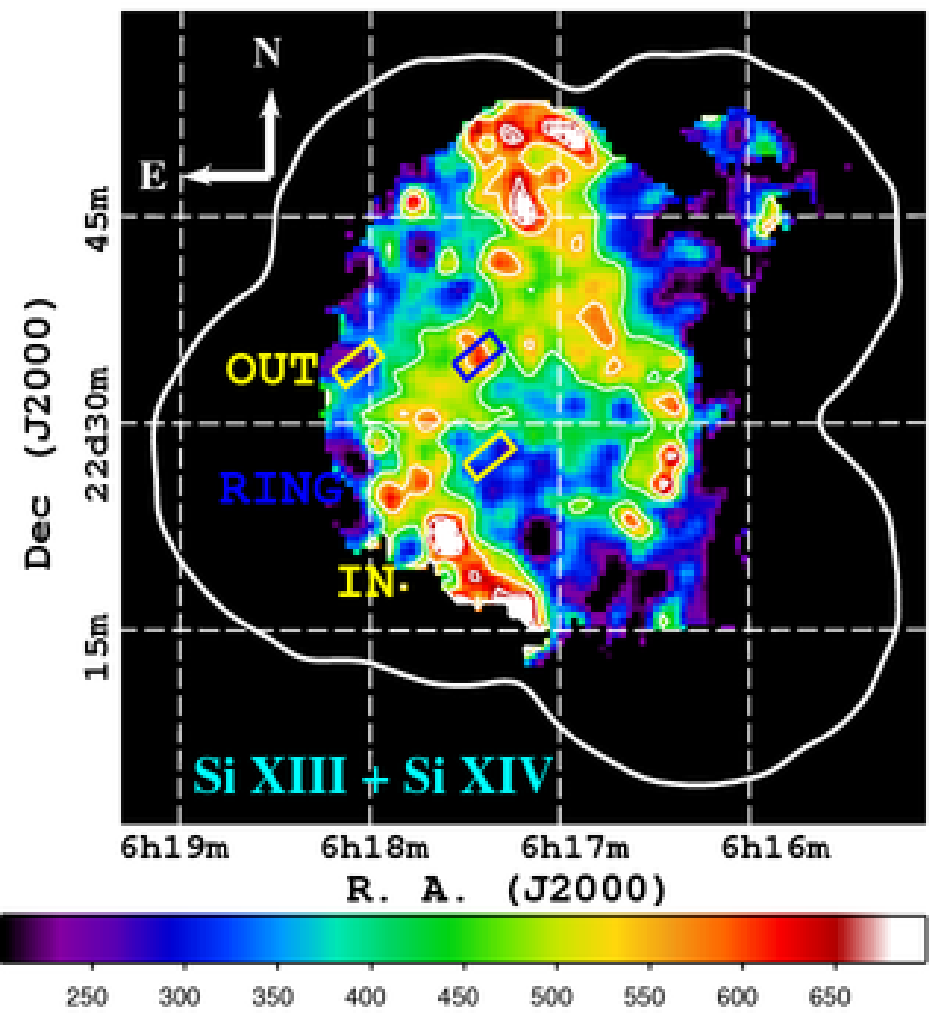,width=\columnwidth}}
     }	
\centerline{} 
 
\centerline{\hbox{
    \psfig{figure=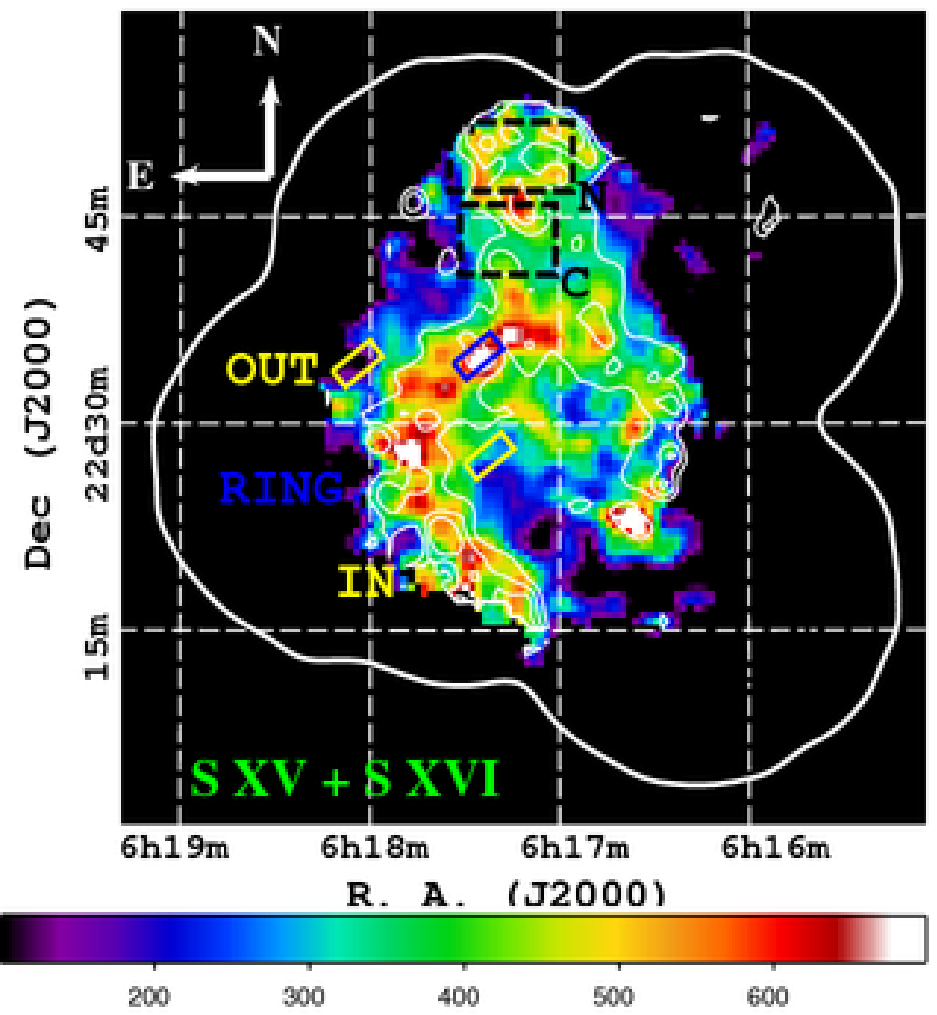,width=\columnwidth}}	 
    }

\caption{Equivalent width
images of Si ($top\ panel$) and S lines ($bottom\ panel$). 
Units are eV.
Noisy regions were masked, as described in the text.
The thin white line mark the FOV.
Contour levels corresponds to 450, 550, 650 eV
in the Si EW image.
We also overlaid the three regions (named IN, RING, and OUT)
selected for spectral analysis.
 The dashed boxes overlaid on the S EW map show the ``North''
and ``Center'' regions, the same of \citet{kawasaki02}, 
and discussed on \S\ref{sec:overion}.} 

\label{fig:ewmap}
\end{figure}

\begin{figure}
\includegraphics[scale=0.36, angle=270]{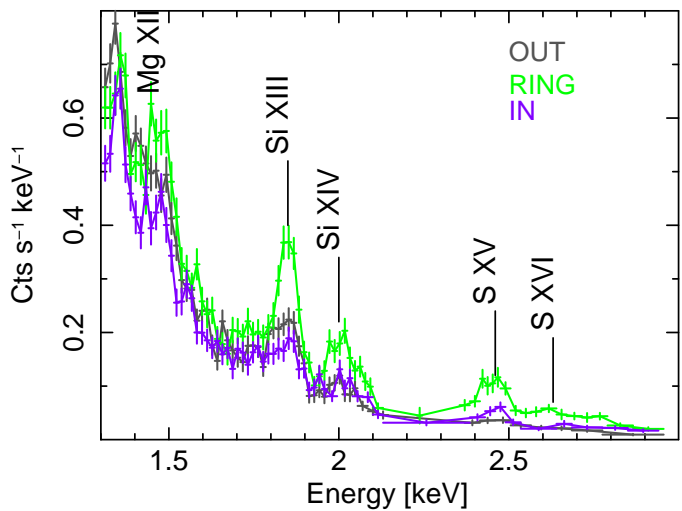}

\caption{A comparison of the three X-ray spectra at high energies,
clearly showing brightest lines in the ``ring'' region respect with
the external (``out'') and the internal (``in'') regions. }
\label{fig:spectra}
\end{figure}

Once an image in each energy band was generated, 
we estimated the continuum emission in the line band (see e.g. \citealt{cassam04}).
In each pixel we computed the underlying continuum as a linear combination of the
source emission in adjacent continuum bands:

\begin{equation}\label{I_c}
\hspace{2.5cm}I_{c}= \lambda \: \alpha_{c}^{-} \: I_{c}^{-}\: + \: (1-\lambda) \:
\alpha_{c}^{+} \: I_{c}^{+}
\end{equation}

where $I_c$ is the resultant continuum image, and $I_c^{\pm}$ are the
background subtracted images in the high and low continuum bands, respectively. 
The normalization coefficients, $\alpha_{c}^{\pm}$, were
determined by the ratio of the model predicted counts and the source
measured counts in the continuum bands, as in \citet{cassam04}. The
weighting factor $\lambda$ was tuned so to minimize statistical fluctuations.

Narrow band images were created with a bin size of 30\arcsec\ to provide at
least 10 counts even in the fainter SNR regions. 

In order to generate EW images, the
image in the line band was background and continuum subtracted and
then divided by the estimated continuum image~$I_c$. 
Since this operation amplified statistical fluctuations, 
we previously applied an adaptive smoothing with 30\arcsec$\leq\sigma\leq$120\arcsec
and a signal to noise ratio equal to 5. 
We finally set the EW to zero in those pixels with a very low estimated
continuum ($<$5\% than the maximum value, see e.g. \citealt{park03b}).

{Si and S EW images are presented in Fig.~\ref{fig:ewmap},
revealing an intriguing circular feature, 10\arcmin in radius, 
in the southern part of the SNR.}
For a morphological comparison we superimposed Si EW contours on the images,
corresponding to 450, 550 and 650 eV.
The two images look very similar.
The contours of high Si EW match well with the bright S EW features.
In particular, we did not find any strong evidence of elemental stratification. 
Both the EW maps show similar features to the median energy map: 
an elongated bright structure in the NE quadrant and a circular ring 
around the PWN region. A rather similar arc-like feature has been observed in
the Si EW map of the SNR~N49 \citep{park03a} and interpreted as the boundary of
hot plasma, re-heated by the reverse shock. The Si arc in N49 is not around the
central compact object but, as in the case of IC~443, it is located near the site of
interaction with a molecular cloud complex, strengthening the hypothesis
of an intense reverse shock, propagating back into the remnant.

The great advantage of EW images is that they 
are not significantly affected by emission measure variations.
However they do also depend on the plasma temperature
and ionization age. 
In order to disentangle multiple effects,
we proceeded with direct spatially resolved spectral analysis,
using the median photon energy and the EW images as
guiding criteria of region selection.

\subsection{Spectral Analysis}\label{sec:spec}

Our aim was to confirm the bright X-ray features visible in Fig.~\ref{fig:q50}
and Fig.~\ref{fig:ewmap}, and give a quantitative characterization of their spectra.
We extracted three EPIC MOS spectra, selected as 
representative on the basis of the median energy map and the EW images.
We selected homogeneous regions, marked in Fig.~\ref{fig:ewmap},
characterized by small fluctuations ($<<$5\%) in the median photon energy and in the line EWs.
Two of them, named ``in'' and ``out'', have low median energies,
low EWs values, and lie inside and outside the ring structure, respectively.
The third region, named ``ring'' and belonging to the ring structure, 
is particularly bright both in the median energy and in the EW images.

Source and background events list were corrected for vignetting effects
with the task \texttt{evigweight} of the SAS package, and then spectra were extracted.
We used the relevant on-axis ancillary response file and 
response matrices generated with the SAS task \texttt{rmfgen}.
Spectra were grouped to a minimum of 25 counts per bin and
the $\chi^2$ statistics was used. 
Fig.~\ref{fig:spectra} show the three spectra 
at high energies, where striking differences in the 
properties of the hot X-ray emitting plasma are clearly visible. 
Prominent Mg, Si, and S lines characterize the  ``ring'' spectrum,
whilst line emission is much fainter in the other two regions. 
Fig.~\ref{fig:spectra} nicely confirms that the identified features 
correspond to different physical conditions of the emitting plasma.

\begin{figure}[!t]
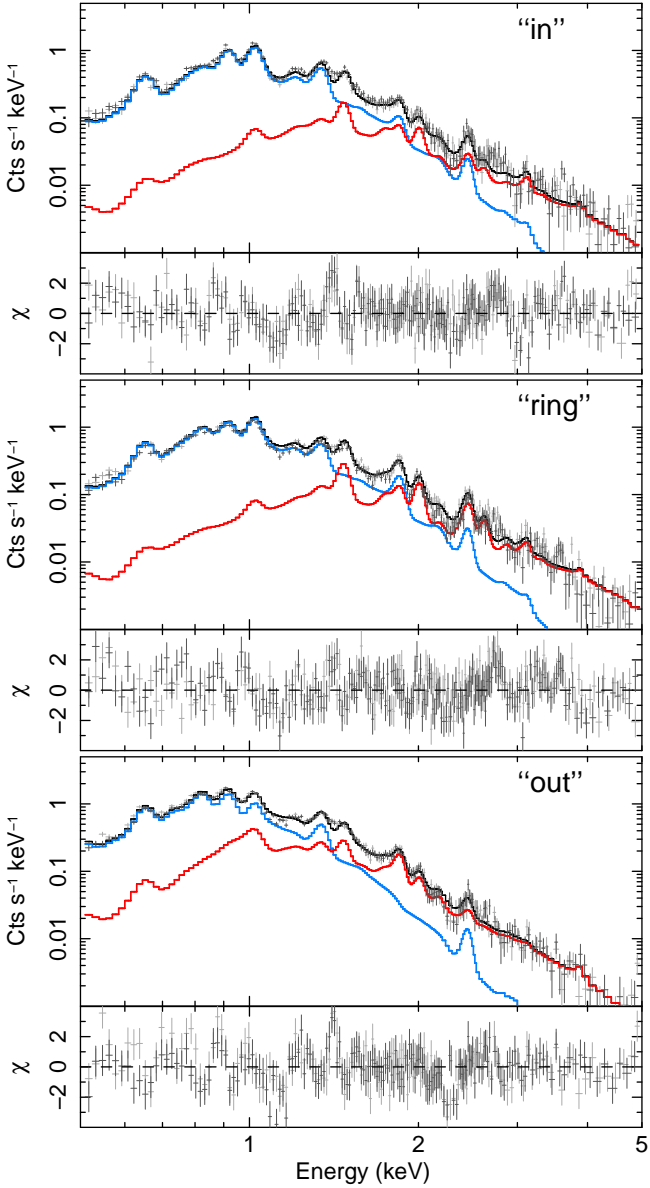

\includegraphics[scale=0.35, angle=270]{f5a.ps} 
\includegraphics[scale=0.35, angle=270]{f5b.ps} 
\includegraphics[scale=0.35, angle=270]{f5c.ps} 
\caption{EPIC MOS spectra of IC~443, selected as representative of the hot X-ray 
 plasma properties, and their best-fit models overlaid. 
 Panels a, b, and c correspond to regions ``in'', ``ring'', and ``out'' 
 in Fig.~\ref{fig:ewmap}, respectively. 
 The lower panel in each plot shows the residuals from the best-fit model.} 
\label{fig:bfit}
\end{figure}

\input{tab2.tex}

We modeled the spectra over the 0.5--5.0~keV energy band
with a NEI plus a CIE thermal component
(VPSHOCK+VMEKAL in \textsc{xspec}), as in \citet{troja06}. 
MOS1 and MOS2 spectra were simultaneously fitted,
leaving the normalization between the two instruments as a free parameter.
We included a 5\% systematic error term 
accounting for the uncertainties in the calibration\footnote{
The current EPIC MOS status of calibration is reported at
http://xmm.vilspa.esa.es/docs/documents/CAL-TN-0018.pdf}. 
The absorbing column \nh\ was estimated by using the median photon energy
in the soft band (cf. \citealt{troja06}, Fig.~6), and held fixed at the
derived value.

Spectra and best-fit models are shown in Fig.~\ref{fig:bfit}.
Table~\ref{tab:spec} reports the spectral properties 
(median photon energy E$_{50}$, Si and S EWs), 
and the best-fit spectral parameters for each region.

\subsubsection{Overionization}\label{sec:overion}

Taking advantage of the large \xmm\ effective area at high energies, 
we checked the overionization claim made by \citet{kawasaki02,kawasaki05}. 
For this purpose, we extracted two spectra from the ``North'' and ``Center'' regions
(see Fig.~2 of \citealt{kawasaki02}), and performed
the same comparison between the ionization and the electron continuum temperatures.

We selected the 2.2--5.0~keV  energy range and we modeled the spectrum 
with a thermal component (model VMEKAL in \textsc{xspec}), fixing the S abundance to zero. 
Three narrow gaussians (centered at 2.46, 2.62 and 2.88~keV) were used to model the sulfur line emission.
The intensity line ratio of the H-like and He-like ions (\sh/\she) is a function of 
the S ionization temperature, derived using the APED line emissivity database
\citep{aped}.

The absorption column densities were held fixed to the values of 0.60$\times$\e{22}\cm{-2} 
for the ``North'' spectrum and 0.68$\times$\e{22}\cm{-2} for the ``Center'' spectrum, 
as derived from the N$_{\rm H}$/E$_{50}$ calibration plot (Fig.~10 in \citealt{troja06}).
We also tested the absorption column value of  0.74$\times$\e{22}\cm{-2}, 
used by \citet{kawasaki02},  but it does not affect the final result 
as expected at these high energies.

\begin{figure}[!t]
\includegraphics[scale=0.5, viewport=24 0 550 350, clip]{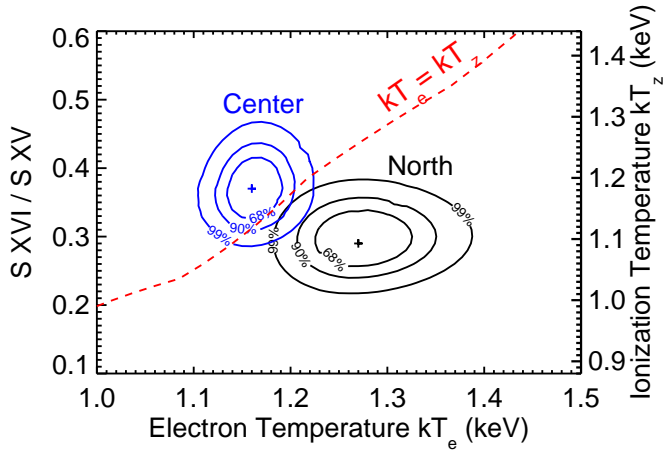}	 

\caption{Confidence level \chisq contours (68\%, 90\%, 99\%)
of the line ratio (\sh/\she) and electron temperature (kT$_{\rm e}$)
for the ``North'' and the ``Center'' region.
On the right Y-axis is reported the ionization temperature kT$_{z}$.
The dashed line indicates the CIE state.}
\label{fig:ktion}
\end{figure}

Fig.~\ref{fig:ktion} shows the 68\%, 90\%, and 99\% \chisq\ contours
in the parameters space \sh/\she -- kT$_{\rm e}$. 
The CIE is represented by the dashed line, tracing the condition kT$_{\rm e}$=kT$_{\rm z}$.
In the ``North'' region we found evidence of an underionized plasma, 
in stark contrast with the findings of \citet{kawasaki02};
in the ``Center'' region the derived ionization temperature is slightly higher
than the the electron temperature,though consistent with it at the 90\% confidence level.
Therefore, 
a detailed analysis of the ionization state does not directly support the presence of 
an overionized plasma in the ``North'' region, and shows only a marginal (1~$\sigma$) evidence
of overionization in the innermost ``Center'' region.

\section{Discussion}\label{sec:discussion}
\subsection{Presence of hot metal-rich plasma in IC~443}\label{sec:plasma}

Spatially-resolved spectral analysis confirms that the structures,
revealed by the median photon energy and the EW maps 
(see \S\ref{sec:median} and \S\ref{sec:ew}, respectively), 
are associated with a high temperature, high metallicity X-ray emitting plasma. 

Our results, listed in Tab.~\ref{tab:spec}, show that the cold plasma is characterized 
by a temperature of 0.3-0.4~keV, a high ionization timescale ($>$\e{12}\,\cm{-3}\,s), 
and solar or undersolar abundances, as expected if the emission arises from the shocked ISM.
The hot plasma is fully equilibrated. It has lower metal abundances 
in the ``in'' and ``out'' regions,
in good agreement with previous findings for the hot component \citep{kawasaki02,petre88}. 
The temperature T$_h$ and the abundances of Mg, Si, and S in the ``ring'' region
are much higher, more similar to regions 2, 6, and 9 analyzed in \citet{troja06}.
As shown in Fig.~\ref{fig:q50} and Fig.~\ref{fig:ewmap}, in the NE the hottest plasma appears concentrated in the innermost regions,
whilst in the south it display a ring-shaped spatial distribution.
Our spectral results fairly agrees with this picture, 
detecting a high temperature and a substantial metal-enrichment 
in correspondence to the ``ring'' region.
The measured overabundant Mg, Si and S are suggestive of SN ejecta emission.

IC~443 is a middle-aged remnant, thus the presence of ejecta at its evolutive stage
is not expected. Models predict that a great part of the ejecta thermalized with the 
surrounding medium, being no longer visible. 
However, \xmm\ and \emph{Chandra} high-resolution observations 
show a growing number of middle-aged remnants ($\sim$10$^4$~yr)
where metal rich ejecta are still detectable and significantly
contribute to the X-ray emission, as in  DEML71 \citep{hughes03}, 
N49B \citep{park03a}, and 0103-72.6 \citep{park03b}.
Moreover, there are several cases among MM SNRs in which 
a soft component, associated with shocked ISM emission,
and a hot metal enriched component have been resolved.  
For instance, \citet{mavroma04} described W63 thermal emission with a 
two temperatures plasma (kT$_1$$\sim$0.2~keV and kT$_2$$\sim$0.6~keV),
showing overabundant Mg, Si and Fe in a bright bar-like region near 
the geometric center of the SNR. 
\citet{lazendic06} resolved a two temperatures thermal plasma
in the inner regions of CTB~1, a prototypical MM SNR, 
and they showed that the hard component (kT$\sim$0.8~keV) 
requires an oversolar Mg abundance.

\citet{kawasaki02,kawasaki05} addressed the issue of overionization
in the hot plasma of MM SNRs, finding that in 2 objects
(namely IC~443 and W49B), out of a sample of 6, some ions 
are indeed overionized.
Such ionization state of the plasma has many important implications, 
since it would indicate rapidly cooling plasma due, for instance, 
to thermal conduction. 
A thermally conductive model has been invoked to explain 
the centrally peaked morphology of MM SNRs \citep{cox99,shelton99},
and overionization would provide a nice consistency check. 

We investigated whether overionization conditions are consistent with 
our \xmm\ data, performing our check in the same regions 
of IC443 analyzed by \citet{kawasaki02}.
In the ``North'' region, we obtained that overionization is not 
consistent with the results for the S ions. 
In the ``Center'' region, overionization is still possible, 
though not required at the 90\% confidence level, for the same ions.
 We point out that this region is characterized by a wide spread
of EWs in both the Si and the S maps ($\Delta$EW/$<$EW$>$$\sim$40\%), therefore 
the analyzed spectrum originates from chemically inhomogeneous regions.
A check of the ionization state would require more uniform conditions.
For instance, \citet{miceli06} found that in the central core of the SNR W49B
there is no compelling evidence of overionization.
Unfortunately, our attempt to analyze smaller but more homogeneous regions
was affected by the low statistics, and did not significatively
constrain the spectral parameters.

\subsection{Constraints on the SN progenitor}\label{sec:sn}

We presented in \S\,\ref{sec:median}-\ref{sec:ew} high resolution maps of some properties
of the hot plasma (median photon energy, and EWs of Si and S lines).
We further investigated and quantified the properties of the X-ray emitting plasma
through a spatially-resolved spectral analysis (\S\,\ref{sec:spec}).
Those results allowed us to infer the spatial distribution 
of the stellar ejecta,
and to estimate their abundances pattern and masses in the ring structure, 
which are strong clues to the nature
of the SN progenitor \citep[e.g.][]{rakowski06,miceli06,park04}, 
and the age of the remnant (see \S\,\ref{sec:age}).

\input{tab3.tex}

Tab.~\ref{tab:abund} reports the best fit metal abundances 
of the ``ring'' region  relative to Si (col.~1), the abundances
ratios expected for core-collapse SNe with different progenitor masses
(col.~2-4; \citealt{ww95}), and for Type~Ia SNe  according to 
different deflagration models (col.~5-6; \citealt{rakowski06,badenes03}).
 Bearing in mind the uncertainty in the predictions of nucleosynthesis yields,
our comparison is aimed at discriminating between Type Ia and Type II SN explosions.
Any further inference, for example on the mass of the progenitor, 
would be highly dependent on the adopted model.

The very low Fe abundance can not meaningfully constrain the progenitor explosion,
and it can be explained if the innermost ejecta layers have not been shocked yet. 
The high Mg abundance allowed us to disfavor a Type Ia SN,
which primarily produces Fe-group elements and a negligible fraction
of lighter metals. A Type~II SN  has been previously suggested as the remnant progenitor,
only on the basis of its location in a star-forming region and the putative association
with a PWN \citep{gaensler06,bocchino01, olbert01}. 
Our results further support and strengthen a core-collapse origin for IC~443.

In order to estimate the total mass of the ejecta,
we assumed that the ejecta are confined within a shell of radius 
$R_{ej}$$\sim$4.5~pc, and width $\Delta$$r$$\sim$1~pc,
for a total emission volume V$_{shell}$$\sim$7.7\ee{57}\,\cm{3}.
Hence the hot plasma fills the entire volume V$_{\rm shell}$ (filling factor $f_h$=1).
The above assumption might not be valid for the Mg ejecta, 
since it is based on the EW maps of Si and S line complexes.
Such high-Z elements are thought to be synthesized
by explosive O- or Si-burning in the core of a massive star,
while lighter metals mostly reside in the outer layers.
If the stellar stratification persists after the explosion,
core-collapse nucleosynthesis models predict the Mg ejecta 
to be preferentially located outside the Si shell
\citep[cf. Fig.~1 of][]{thielemann96}, thus occupying a larger volume.
The possibility of mixing or overturning of the ejecta layers 
\citep{willingale02,hwang00} add further uncertainty. 
Therefore the estimated mass of Mg should be taken with care.

For the case where the electrons are primarily from
H/He ($n_e$=1.2\,$n_H$), 
we derived a hot component gas density of $n_H$=(0.62$\pm$0.10)\,\cm{-3}
in our ``ring'' region, which intercepts a line of sight of $\sim$2~pc.
By assuming the same line of sight for the inner and outer regions, 
we estimated a density of $n_H$=(0.60$\pm$0.10)\,\cm{-3}, and 
$n_H$=(0.96$\pm$0.16)\,\cm{-3}, respectively.

We extrapolated the density derived in the ``ring'' region 
to the whole shell of ejecta, thus estimating the total mass
of hot metal-rich plasma, M$_{\rm tot}$$\sim$4\,M$_{\odot}$.
We calculated a lower limit on the mass of ejecta M$_Z$ 
as
${\rm M}_Z/{\rm M}_{\rm tot}\geq
{\rm A}_Z\times({\rm M}_Z/{\rm M}_{\rm tot})_{\odot}$ 
\citep[e.g.][]{lazendic05},
where M$_{\rm tot}$ is the total mass of the hot X-ray-emitting plasma, 
A$_Z$ is the measured abundance for the element $Z$, 
and $({\rm M}_Z/{\rm M}_{\rm tot})_{\odot}$ 
is the standard solar mass fraction
(6.4\ee{-4} for Mg, 7.0\ee{-4} for Si, 
and 3.6\ee{-4} for S; \citealt{anders89}).
We obtained: M$_{\rm Mg}$$\gtrsim$0.013\,\msun,
M$_{\rm Si}$$\gtrsim$0.007\,\msun,
M$_{\rm S}$$\gtrsim$0.003\,\msun.

An upper limit on M$_Z$ was derived by the assumption of ``pure'' $Z$-ejecta,
i.e. stellar fragments in which all the electrons
are provided by the ionized specie $Z$, $n_e$$\sim$K$_Z$$n_Z$. 
This is a crude approximation of the ejecta mass,
but not unrealistic as this kind of objects have already been observed
in other SNRs, as the Si-rich knots in Cas~A \citep{laming03}.
We considered the contribution from both the ionization states
of Si and S (K$_{\rm Si}$=12.5, K$_{\rm S}$=14.5), but
only the highest ionization state of Mg (K$_{\rm Mg}$=11),
as results from the best fit model (see Fig.~\ref{fig:bfit}, middle panel).
We obtained: M$_{\rm Mg}$$\lesssim$0.55\,\msun,
M$_{\rm Si}$$\lesssim$0.30\,\msun,
M$_{\rm S}$$\lesssim$0.15\,\msun.

It is worth noting that according to \citet{ww95} the predicted masses 
of Mg, Si and S ejecta  lie in the ranges between 0.01--0.4\,\msun,
0.02--0.4\,\msun, and 0.01--0.2\,\msun, respectively. 
These values well agree with the ranges derived from our analysis, 
which can be considered meaningful constraints on the ejecta masses
in the hot shell.

\begin{figure}[!t]
\centerline{\hbox{
     \psfig{figure=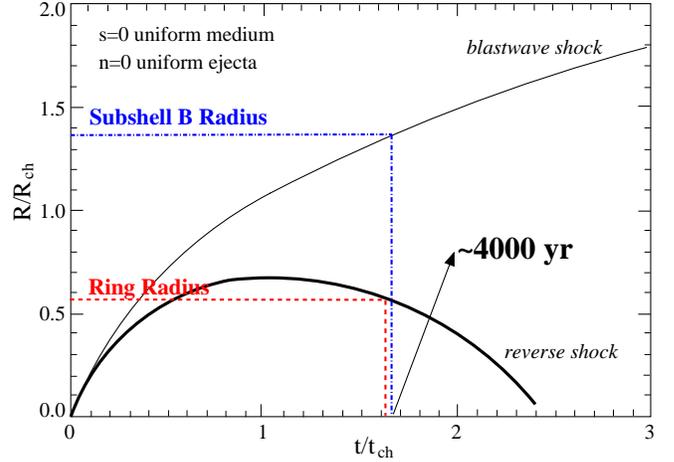,width=0.95\columnwidth}	 
     }}
\centerline{}     
\caption{Shock positions for blast-wave (thin solid line) and reverse (thick solid line)
shocks in remnants of uniform ejecta and evolving in an uniform medium \citep{tm99}.
A comparison with the observed values for IC~443 (r$_1$$\sim$11 pc for the 
subshell B, and r$_2$$\sim$4.5 pc for the ring of hot gas, tracing the forward 
and reverse shock location, respectively) gives a SNR age of 4\ee{3}\,yr.}
\label{fig:age}
\end{figure}

\subsection{Age of IC~443}\label{sec:age}

We examined the case of a SNR expanding in an 
uniform ambient medium ($s$=0) and with an uniform ejecta distribution ($n$=0).
Similar results are found when considering a power law ejecta distribution ($n$=2).
We took our estimate of the pre-shock density in the eastern regions (subshell B), 
n$_0$$\sim$0.25\,\cm{-3},
and the mass of the hot plasma calculated in \S\ref{sec:sn}.
For typical SN energies ($\sim$\e{51}\,erg), the length scale and the time scale 
of the \citet{tm99} solution are $R_{ch}\sim$8~pc and $t_{ch}\sim$2.3\ee{3} yrs, respectively.

Fig.~\ref{fig:age} show the predictions of \citet{tm99} compared with the observed parameters
of IC~443: a forward shock radius of 11~pc (shell B radius) 
and a reverse shock radius of 4.5~pc (ring radius),
correspond to an evolutive age of $\sim$4000 yr, much younger than some previous estimates.
In particular, \citet{cheva99} derived an age of 3\ee{4}\,yr by assuming that 
the remnant has been evolving in a very dense environment ($\sim$15\,\cm{-3}) 
since its early stages. Our result is instead more similar to the age 
estimated by \citet{petre88}, who considered an exponential density distribution 
for the circumstellar environment.
The analysis of the soft X-ray emission from IC~443,
and the derived density profile \citep[cf.][Fig.~7]{troja06}
suggest that the blast wave propagated in a much lower density medium,
and only recently has encountered a sharp density gradient in the NE
(subshell A), impacting onto a  neutral hydrogen cloud \citep{rho01,denoyer78}.

As derived in \S\ref{sec:sn}, the remnant likely originated
from the collapse of a massive star. Hence, during the progenitor lifetime
strong stellar winds could have blown away the surrounding molecular clouds,
forming a low density cavity, where the remnant is now evolving.
\citet{braun86} first argued that the multiple shell morphology of IC~443
might be the result of such pre-SN interaction, and that 
the blast wave is impacting on the pre-existing wind-bubble walls.
A detailed kinematical study of the optical filaments in IC~443,
carried out by \citet{meaburn90}, gave further support to the 
scenario proposed by \citet{braun86}.
The authors detected high velocity features along their slit Pos. 9, at the edge between subshell A and B, 
which can be easily explained if the emitting material, accelerated by the passage of the shock, is seen edge-on. 
The resultant geometry  of the optical filaments is fully consistent with an SNR evolving  
within pre-existing interconnected cavities (cf. \citealt{meaburn90}, Fig.~5).

 Since the medium surrounding massive stars is considerably modified by  
pre-SN winds \citep[e.g.][]{chu03,dwarkadas05}, we verified whether our estimate 
of the age of IC443 significantly depends on the assumption of a uniform 
ambient medium. For this purpose, we modeled the evolution of a SNR in an 
isothermal medium with a power-law density profile $\rho$\,$\propto$\,$r^{-2}$, 
by performing a numerical hydrodynamic simulation. 
We used the FLASH code (Fryxell 2000) and we set up a spherically symmetric 
explosion of 3\,\msun\ of ejecta with a total energy of $10^{51}$ erg. 
The initial velocity profile of the ejecta increases linearly with their 
distance from the center, and their density profile decreases exponentially 
(in agreement with \citealt{wang02}). 
In this case $\rho_s$$=$$\dot M_w/(4 \pi v_w)$$\sim$$1.76\times 10^{14}$\,g\,cm$^{-1}$, 
where $\dot M_w$ is the mass-loss rate of the wind and $v_w$
is its speed, and both are assumed to be constant. 
The location of the forward and reverse shock fronts obtained in the 
simulation suggests a slightly younger age ($\sim$3000~yr). 
Our basic assumption, i. e. that the radial separation between the forward and the reverse shock 
is the result of the SNR evolutive stage, does not hold if the forward shock in the NW has been strongly 
decelerated in the cavity walls. Modifications of the shock velocity, due to the interaction of the blast 
wave with the confining walls, led to systematically underestimate the SNR age. In that case, the inferred value
of 4000~yr only provides a lower limit.
However, in such a scenario, a close correlation between the two radii,
as shown in Fig.~\ref{fig:age}, is not expected, 
as the forward shock radius would mainly depend on the properties of the cavity wall.
We therefore conclude that the morphology of the X-ray emitting plasma in IC~443
robustly suggests an age $<$\e{4} yr.

\subsubsection{Implications for the PWN}

A detailed study of the morphological and spectral properties of the PWN, 
and its central neutron star (NS; J0617+2221) has been presented by \citet{gaensler06}.
They inferred a blackbody temperature kT$^{\infty}$$\sim$100 eV for the NS, 
and argued that the properties of the thermal radiation from the NS
are consistent with the \citet{cheva99} age of 30,000 yrs.
By using the theoretical cooling curve of \citet{kaminker02},
we checked that our new estimate of the SNR age and the temperatures range
measured by \citet{gaensler06} are still consistent with the cooling of a 
1.35--1.45 \msun\ NS with a 1p proton superfluid core. 
Thus, the present knowledge of NS cooling curves does not allow us to
put tighter constraints on the NS age, that would strengthen the association 
between the remnant and the PWN.
The NS temperature of \citet{gaensler06}
and a putative age $<$\e{4}\,yr are in agreement with observational results
from other young NSs: PSR~J1119-6127 ($\sim$1.6 kyr)
RX~J0822-43 ($\sim$3.7 kyr) 1E 1207-52, and PSR J1357-6429 ($\sim$7 kyr;
\citealt{zavlin07a,zavlin07b, kaminker02}).

The location of the ring of hot metal-rich plasma in our Fig.~2 and Fig.~4,
just around the PWN, is an additional argument in support of the physical 
connection between the PWN and IC~443.
On the plane of the image (see e.g. Fig.~2) only the south side of the PWN 
seems to be in contact with the reverse shock, suggesting
that the PWN may be in the reverse shock or subsonic expansion stage,
but not in the high supersonic stage \citep{swaluw04}.
This is consistent with an age of few thousand years, and in rough agreement
with the mildly supersonic Mach value found by \citet{gaensler06}.

The orientation of the cometary tail and the position of the PWN 
cannot be straightforwardly explained. 
This mismatch is still an open issue, but it does not precludes the association between the PWN, 
the ring and the SNR \citep[see also][]{gaensler06}. For instance, a similar apparent offset between the stellar remnant,
the forward, and the reverse shock geometric centers has been already observed in Cas~A \citep{gotthelf01}.
A possibility is that strong reverse shocks, 
originated by the interaction with the surrounding clouds, 
skewed the orientation of the nebula. 
An asymmetric SN explosion or the presence of strong density gradients across the remnant
offer a valid explanation to the observed offsets \citep{gotthelf01,reed95}.
Another possible explanation is that the SN ejecta have been decelerated by the southern ridge of molecular material,
and, as a consequence, the geometric center of the ring appears shifted toward the north direction.

\section{Conclusions}\label{sec:end}
We analyzed the hard (1.4-5.0~keV) X-ray emission of the supernova remnant IC~443,
as observed by a set of public archive \xmm\ observations.
Unlike the soft X-ray emission, which traces directly the interaction with the surrounding clouds
\citep{troja06}, the hard X-ray emission has a centrally-peaked morphology,
reminiscent of the subclass of mixed morphology SNRs.
 The presence of a ring-shaped feature, which encircles the PWN, has been revealed 
in the equivalent width maps of Si and S lines (Fig. 3).

A spatially-resolved spectral analysis confirms that this structure is associated to hot
metal rich plasma, whose abundances are consistent with a core-collapse SN origin,
but not with an exploding white dwarf. 
We argue that the ring indicates the location of an ejecta layer heated by the reverse shock,
and that the reverse shock has partially hit the PWN.
Dynamically, the location of the ejecta ring suggests an age of $\sim$4,000 yr for IC~443,
an order of magnitude lower than  some previous estimates, based on expansion in a very dense environment.
A younger age is indeed more consistent with an SNR expansion within wind-blown bubble shells,
surrounded by dense clouds, as first suggested by \citet{braun86}, and recently
considered by \citet{troja06}.

 We detect only marginal evidence of plasma overionization in the NE part of the remnant.
The bright hard \mbox{1.4--5.0~keV} X-ray emission in the NE region of IC~443 also show high temperatures and high metal abundances, as shown in \citet{troja06}.
More detailed results from this region will be reported in a future work.

\section*{Acknowledgments}
We thank the anonymous referee for his/her insightful comments on this paper.
We also thank J. Raymond and R. Chevalier for useful discussions about 
the optical and the X-ray emission of IC~443, and Richard Owen for
his kind help.

FB and MM acknowledge financial contribution from the contract
ASI-INAF I/203/05. 
The software used in this work was in part developed by the DOE-supported 
ASC / Alliance Center for Astrophysical Thermonuclear Flashes at the 
University of Chicago.
This work makes use of results produced by the PI2S2 Project managed by the
Consorzio COMETA, a project co-funded by the Italian Ministry of University
and Research (MIUR) within the Piano Operativo Nazionale "Ricerca Scientifica,
Sviluppo Tecnologico, Alta Formazione" (PON 2000-2006). More information
is available at http://www.pi2s2.it and http://www.consorzio-cometa.it.

\bibliography{myreferences}

\end{document}

%% file: tab1.tex

\begin{table}[t!]
\begin{center}
\caption{Energy bands selected for the equivalent width images.}
\label{tab:bands}
\begin{tabular}{cccc}
\hline\hline
Elements  & Lines & Low continuum &  High continuum\\
          & (eV)     & (eV)         & (eV)  \\
\hline
 Si$\dotfill$ 	& 1790-2060
    		& 1630-1690
    		& 2180-2320 \\
 S$\dotfill$  	& 2360-2680
    		& 2180-2320
    		& 3350-3520 \\
\hline
\end{tabular}
\end{center}
\end{table}


%% file: tab2.tex

\begin{table}[!t]
\begin{center}
\caption{ Spectral parameters of 
the three analysed regions.} 
\label{tab:spec}
\begin{tabular}{lccc}
\hline
\hline
{Parameters \ \ \ \ \ \ \ \ \ \ \ \ \ \ \ \ \ \ } & IN & RING & OUT \\
\hline
\vspace{0.15cm} E$_{50}$ (keV) $\dotfill$      
            & 1.70
            & 1.78
            & 1.65 \\
\vspace{0.15cm} Si EW (eV) $\dotfill$        
            &  310
            &  550
            &  290 \\
\vspace{0.15cm} S EW (eV) $\dotfill$        
            &  280 
            &  690
            &  130 \\	 	        
\hline
\multicolumn{4}{c}{Cold component}\\ 
\hline
\vspace{0.15cm} \nh\ (\e{22}\cm{-2}) $\dotfill$      
            & [0.61]
            & [0.58]
            & [0.50] \\
\vspace{0.15cm} kT$_s$ (keV) $\dotfill$  
            & 0.44$^{+0.04}_{-0.03}$ 
	    & 0.43$\pm$0.02
            & 0.32$^{+0.03}_{-0.02}$  \\
\vspace{0.15cm} $\tau$ (\e{12}\cms) $\dotfill$      
            & 1.7$^{+0.04}_{-0.06}$
            & 1.8$\pm$0.05
            & 2.5$^{+0.5}_{-1.0}$        \\
\vspace{0.15cm} O/O$_{\odot}$ $\dotfill$       
            & 0.78$^{+0.16}_{-0.2}$
            & 0.68$^{+0.2}_{-0.08}$   
            & 0.30$\pm$0.10      \\
\vspace{0.15cm} Ne/Ne$_{\odot}$ $\dotfill$ 
            & 0.91$^{+0.3}_{-0.10}$
            & 0.81$^{+0.12}_{-0.3}$
            & 0.35$^{+0.10}_{-0.05}$  \\    
\vspace{0.15cm} Mg/Mg$_{\odot}$ $\dotfill$      
            & 0.45$^{+0.09}_{-0.05}$
            & 0.35$\pm$0.05 
            & 0.35$\pm$0.05 \\
\vspace{0.15cm} Si/Si$_{\odot}$ $\dotfill$ 
            & 0.26$^{+0.06}_{-0.11}$
            & 0.50$^{+0.10}_{-0.15}$
            & $<$0.15  \\    
\vspace{0.15cm} Fe/Fe$_{\odot}$ $\dotfill$     
            & 0.10
            & 0.20$\pm$0.03
            & 0.22$\pm$0.05 \\	
\vspace{0.15cm} EM$_s$\,(\e{56}\,\cm{-3})$\dotfill$ 
            & 4.1
            & 5.7
            & 8.7 \\	        
\hline
\multicolumn{4}{c}{Hot component}\\ 
\hline
\vspace{0.15cm} kT$_h$ (keV) $\dotfill$  
            & 1.35$^{+0.4}_{-0.19}$
            & 1.45$^{+0.08}_{-0.15}$
            & 0.86$\pm$0.05    \\
\vspace{0.15cm} Mg/Mg$_{\odot}$ $\dotfill$       
            & 2.6$^{+3}_{-1.1}$
            & 5.0$^{+3}_{-0.9}$
            & [1] \\
\vspace{0.15cm} Si/Si$_{\odot}$ $\dotfill$       
            & 1.1$\pm$0.3
            & 2.4$\pm$0.7
            & 0.67$^{+0.19}_{-0.11}$\\
\vspace{0.15cm} S/S$_{\odot}$ $\dotfill$       
            & 0.7$\pm$0.5
            & 2.3$^{+0.6}_{-0.7}$
            & 0.22$^{+0.12}_{-0.11}$   \\
\vspace{0.15cm} Fe/Fe$_{\odot}$ $\dotfill$      
            & $<$0.13
            & $<$0.03 
            & $<$0.23 \\
\vspace{0.15cm} EM$_h$\,(\e{56}\,\cm{-3})$\dotfill$ 
            & 1.1
            & 0.4
            & 0.4 \\	    
$\chi^2$/dof $\dotfill$      
            &  420/324
            &  497/343
            &  412/302 \\
\hline

\end{tabular}
\end{center}
\end{table}

%% file: tab3.tex

\begin{table}[!bh]
\begin{center}
\caption{  
Relative abundances in the ring region: [X/Si]/[X/Si]$_{\odot}$
}

\label{tab:abund}
\begin{tabular*}{0.95\columnwidth}{@{\extracolsep{-4pt}}lcccccc}
\hline
\hline
{Ratio\ \ \ \ \ \ \ } & IC~443 & 20 M$_{\odot}$$^a$
& 25 M$_{\odot}$$^a$ & 30 M$_{\odot}$$^a$ & PDDe$^b$ & DDTe$^b$\\
\hline
\vspace{0.15cm} 
Mg/Si\dotfill  & 2.1$^{+1.4}_{-0.7}$ & 0.16 &  0.45 &  1.79  & 0.0017 & 0.025 \\
\vspace{0.15cm}
S/Si\dotfill   & 1.0$\pm$0.4 & 1.28 & 0.96 & 0.24 & 1.5 & 1.4  \\
Fe/Si\dotfill   & $<$0.018 & 0.88 & 0.13 & 0.15 & 0.89 & 0.91  \\
\hline
\end{tabular*}
\end{center}

$^a$\ \citet{ww95}.
$^b$\ Pulsating delayed detonation (PDDe)
and delayed detonation (DDTe) models from \citet{badenes03}.

\end{table}


%% file: ejecta.bbl
\begin{thebibliography}{68}
\expandafter\ifx\csname natexlab\endcsname\relax\def\natexlab#1{#1}\fi

\bibitem[{{Albert} {et~al.}(2007){Albert}, {Aliu}, {Anderhub}, {Antoranz},
  {Armada}, {Baixeras}, {Barrio}, {Bartko}, {Bastieri}, {Becker}, {Bednarek},
  {Berger}, {Bigongiari}, {Biland}, {Bock}, {Bordas}, {Bosch-Ramon}, {Bretz},
  {Britvitch}, {Camara}, {Carmona}, {Chilingarian}, {Coarasa}, {Commichau},
  {Contreras}, {Cortina}, {Costado}, {Curtef}, {Danielyan}, {Dazzi}, {De
  Angelis}, {Delgado}, {de los Reyes}, {De Lotto}, {Domingo-Santamar{\'{\i}}a},
  {Dorner}, {Doro}, {Errando}, {Fagiolini}, {Ferenc}, {Fern{\'a}ndez}, {Firpo},
  {Flix}, {Fonseca}, {Font}, {Fuchs}, {Galante}, {Garc{\'{\i}}a-L{\'o}pez},
  {Garczarczyk}, {Gaug}, {Giller}, {Goebel}, {Hakobyan}, {Hayashida},
  {Hengstebeck}, {Herrero}, {H{\"o}hne}, {Hose}, {Hsu}, {Jacon}, {Jogler},
  {Kosyra}, {Kranich}, {Kritzer}, {Laille}, {Lindfors}, {Lombardi}, {Longo},
  {L{\'o}pez}, {L{\'o}pez}, {Lorenz}, {Majumdar}, {Maneva}, {Mannheim},
  {Mansutti}, {Mariotti}, {Mart{\'{\i}}nez}, {Mazin}, {Merck}, {Meucci},
  {Meyer}, {Miranda}, {Mirzoyan}, {Mizobuchi}, {Moralejo}, {Nieto}, {Nilsson},
  {Ninkovic}, {O{\~n}a-Wilhelmi}, {Otte}, {Oya}, {Paneque}, {Panniello},
  {Paoletti}, {Paredes}, {Pasanen}, {Pascoli}, {Pauss}, {Pegna}, {Persic},
  {Peruzzo}, {Piccioli}, {Prandini}, {Puchades}, {Raymers}, {Rhode},
  {Rib{\'o}}, {Rico}, {Rissi}, {Robert}, {R{\"u}gamer}, {Saggion}, {Saito},
  {S{\'a}nchez}, {Sartori}, {Scalzotto}, {Scapin}, {Schmitt}, {Schweizer},
  {Shayduk}, {Shinozaki}, {Shore}, {Sidro}, {Sillanp{\"a}{\"a}}, {Sobczynska},
  {Stamerra}, {Stark}, {Takalo}, {Temnikov}, {Tescaro}, {Teshima}, {Torres},
  {Turini}, {Vankov}, {Vitale}, {Wagner}, {Wibig}, {Wittek}, {Zandanel},
  {Zanin}, \& {Zapatero}}]{albert07}
{Albert}, J., {et~al.} 2007, \apjl, 664, L87

\bibitem[{{Anders} \& {Grevesse}(1989)}]{anders89}
{Anders}, E. \& {Grevesse}, N. 1989, \gca, 53, 197

\bibitem[{{Arnaud}(1996)}]{arnaud96}
{Arnaud}, K.~A. 1996, in ASP Conf. Ser. 101: Astronomical Data Analysis
  Software and Systems V, 17--+

\bibitem[{{Asaoka} \& {Aschenbach}(1994)}]{asaoka94}
{Asaoka}, I. \& {Aschenbach}, B. 1994, \aap, 284, 573

\bibitem[{{Badenes} {et~al.}(2003){Badenes}, {Bravo}, {Borkowski}, \&
  {Dom{\'{\i}}nguez}}]{badenes03}
{Badenes}, C., {et~al.} 2003, \apj, 593, 358

\bibitem[{{Bocchino} \& {Bykov}(2000)}]{bocchino00}
{Bocchino}, F. \& {Bykov}, A.~M. 2000, \aap, 362, L29

\bibitem[{{Bocchino} \& {Bykov}(2001)}]{bocchino01}
{Bocchino}, F. \& {Bykov}, A.~M. 2001, \aap, 376, 248

\bibitem[{{Bocchino} \& {Bykov}(2003)}]{bocchino03b}
{Bocchino}, F. \& {Bykov}, A.~M. 2003, \aap, 400, 203

\bibitem[{{Braun} \& {Strom}(1986)}]{braun86}
{Braun}, R. \& {Strom}, R.~G. 1986, \aap, 164, 193

\bibitem[{{Burton} {et~al.}(1990){Burton}, {Hollenbach}, {Haas}, \&
  {Erickson}}]{burton90}
{Burton}, M.~G., {et~al.} 1990, \apj, 355, 197

\bibitem[{{Bykov} {et~al.}(2005){Bykov}, {Bocchino}, \& {Pavlov}}]{bykov05}
{Bykov}, A.~M., {et~al.} 2005, \apjl, 624, L41

\bibitem[{{Bykov} {et~al.}(2008){Bykov}, {Krassilchtchikov}, {Uvarov},
  {Bloemen}, {Bocchino}, {Dubner}, {Giacani}, \& {Pavlov}}]{bykov08}
{Bykov}, A.~M., {et~al.} 2008, \apj, in press, arXiv: astro-ph/0801.1255, 801

\bibitem[{{Cassam-Chena{\"i}} {et~al.}(2004){Cassam-Chena{\"i}},
  {Decourchelle}, {Ballet}, {Hwang}, {Hughes}, {Petre}, \& {et al.}}]{cassam04}
{Cassam-Chena{\"i}}, G., {et~al.} 2004, \aap, 414, 545

\bibitem[{{Chevalier}(1999)}]{cheva99}
{Chevalier}, R.~A. 1999, \apj, 511, 798

\bibitem[{{Chu} {et~al.}(2003){Chu}, {Gruendl}, \& {Guerrero}}]{chu03}
{Chu}, Y.-H., {et~al.} 2003, in Revista Mexicana de Astronomia y Astrofisica,
  vol. 27, Vol.~15, Revista Mexicana de Astronomia y Astrofisica Conference
  Series, ed. J.~{Arthur} \& W.~J. {Henney}, 62--67

\bibitem[{{Claussen} {et~al.}(1997){Claussen}, {Frail}, {Goss}, \&
  {Gaume}}]{claussen97}
{Claussen}, M.~J., {et~al.} 1997, \apj, 489, 143

\bibitem[{{Cornett} {et~al.}(1977){Cornett}, {Chin}, \& {Knapp}}]{cornett77}
{Cornett}, R.~H., {et~al.} 1977, \aap, 54, 889

\bibitem[{{Cox} {et~al.}(1999){Cox}, {Shelton}, {Maciejewski}, {Smith},
  {Plewa}, {Pawl}, \& {R{\'o}{\.z}yczka}}]{cox99}
{Cox}, D.~P., {et~al.} 1999, \apj, 524, 179

\bibitem[{{Denoyer}(1977)}]{denoyer77}
{Denoyer}, L.~K. 1977, \apj, 212, 416

\bibitem[{{Denoyer}(1978)}]{denoyer78}
{Denoyer}, L.~K. 1978, \mnras, 183, 187

\bibitem[{{Dickel} {et~al.}(1989){Dickel}, {Williamson}, {Mufson}, \&
  {Wood}}]{dickel89}
{Dickel}, J.~R., {et~al.} 1989, \aj, 98, 1363

\bibitem[{{Duin} \& {van der Laan}(1975)}]{duin75}
{Duin}, R.~M. \& {van der Laan}, H. 1975, \aap, 40, 111

\bibitem[{{Dwarkadas}(2005)}]{dwarkadas05}
{Dwarkadas}, V.~V. 2005, \apj, 630, 892

\bibitem[{{Esposito} {et~al.}(1996){Esposito}, {Sreekumar}, {Hunter}, \&
  {Kanbach}}]{esposito96}
{Esposito}, J.~A., {et~al.} 1996, Bulletin of the American Astronomical
  Society, 28, 1346

\bibitem[{{Gaensler} {et~al.}(2006){Gaensler}, {Chatterjee}, {Slane}, {van der
  Swaluw}, {Camilo}, \& {Hughes}}]{gaensler06}
{Gaensler}, B.~M., {et~al.} 2006, \apj, 648, 1037

\bibitem[{{Gotthelf} {et~al.}(2001){Gotthelf}, {Koralesky}, {Rudnick}, {Jones},
  {Hwang}, \& {Petre}}]{gotthelf01}
{Gotthelf}, E.~V., {et~al.} 2001, \apjl, 552, L39

\bibitem[{{Hewitt} {et~al.}(2006){Hewitt}, {Yusef-Zadeh}, {Wardle}, {Roberts},
  \& {Kassim}}]{hewitt06}
{Hewitt}, J.~W., {et~al.} 2006, \apj, 652, 1288

\bibitem[{{Hoffman} {et~al.}(2003){Hoffman}, {Goss}, {Brogan}, {Claussen}, \&
  {Richards}}]{hoffman03}
{Hoffman}, I.~M., {et~al.} 2003, \apj, 583, 272

\bibitem[{{Hughes} {et~al.}(2003){Hughes}, {Ghavamian}, {Rakowski}, \&
  {Slane}}]{hughes03}
{Hughes}, J.~P., {et~al.} 2003, \apjl, 582, L95

\bibitem[{{Hwang} {et~al.}(2000){Hwang}, {Holt}, \& {Petre}}]{hwang00}
{Hwang}, U., {et~al.} 2000, \apjl, 537, L119

\bibitem[{{Kaminker} {et~al.}(2002){Kaminker}, {Yakovlev}, \&
  {Gnedin}}]{kaminker02}
{Kaminker}, A.~D., {et~al.} 2002, \aap, 383, 1076

\bibitem[{{Kawasaki} {et~al.}(2005){Kawasaki}, {Ozaki}, {Nagase}, {Inoue}, \&
  {Petre}}]{kawasaki05}
{Kawasaki}, M., {et~al.} 2005, \apj, 631, 935

\bibitem[{{Kawasaki} {et~al.}(2002){Kawasaki}, {Ozaki}, {Nagase}, {Masai},
  {Ishida}, \& {Petre}}]{kawasaki02}
{Kawasaki}, M.~T., {et~al.} 2002, \apj, 572, 897

\bibitem[{{Laming} \& {Hwang}(2003)}]{laming03}
{Laming}, J.~M. \& {Hwang}, U. 2003, \apj, 597, 347

\bibitem[{{Lampton} {et~al.}(1976){Lampton}, {Margon}, \& {Bowyer}}]{lampton76}
{Lampton}, M., {et~al.} 1976, \apj, 208, 177

\bibitem[{{Lazendic} \& {Slane}(2006)}]{lazendic06}
{Lazendic}, J.~S. \& {Slane}, P.~O. 2006, \apj, 647, 350

\bibitem[{{Lazendic} {et~al.}(2005){Lazendic}, {Slane}, {Hughes}, {Chen}, \&
  {Dame}}]{lazendic05}
{Lazendic}, J.~S., {et~al.} 2005, \apj, 618, 733

\bibitem[{{Leahy}(2004)}]{leahy04}
{Leahy}, D.~A. 2004, \aj, 127, 2277

\bibitem[{{Mavromatakis} {et~al.}(2004){Mavromatakis}, {Aschenbach}, {Boumis},
  \& {Papamastorakis}}]{mavroma04}
{Mavromatakis}, F., {et~al.} 2004, \aap, 415, 1051

\bibitem[{{Meaburn} {et~al.}(1990){Meaburn}, {Whitehead}, {Raymond}, {Clayton},
  \& {Marston}}]{meaburn90}
{Meaburn}, J., {et~al.} 1990, \aap, 227, 191

\bibitem[{{Miceli} {et~al.}(2006){Miceli}, {Decourchelle}, {Ballet},
  {Bocchino}, {Hughes}, {Hwang}, \& {Petre}}]{miceli06}
{Miceli}, M., {et~al.} 2006, \aap, 453, 567

\bibitem[{{Neufeld} \& {Yuan}(2008)}]{neufeld08}
{Neufeld}, D. \& {Yuan}, Y. 2008, \apj, in press,arXiv: astro-ph/0801.2141

\bibitem[{{Olbert} {et~al.}(2001){Olbert}, {Clearfield}, {Williams}, {Keohane},
  \& {Frail}}]{olbert01}
{Olbert}, C.~M., {et~al.} 2001, \apjl, 554, L205

\bibitem[{{Park} {et~al.}(2003{\natexlab{a}}){Park}, {Burrows}, {Garmire},
  {Nousek}, {Hughes}, \& {Williams}}]{park03a}
{Park}, S., {et~al.} 2003{\natexlab{a}}, \apj, 586, 210

\bibitem[{{Park} {et~al.}(2004){Park}, {Hughes}, {Slane}, {Burrows}, {Roming},
  {Nousek}, \& {Garmire}}]{park04}
{Park}, S., {et~al.} 2004, \apjl, 602, L33

\bibitem[{{Park} {et~al.}(2003{\natexlab{b}}){Park}, {Hughes}, {Slane},
  {Burrows}, {Warren}, {Garmire}, \& {Nousek}}]{park03b}
{Park}, S., {et~al.} 2003{\natexlab{b}}, \apjl, 592, L41

\bibitem[{{Petre} {et~al.}(1988){Petre}, {Szymkowiak}, {Seward}, \&
  {Willingale}}]{petre88}
{Petre}, R., {et~al.} 1988, \apj, 335, 215

\bibitem[{{Rakowski} {et~al.}(2006){Rakowski}, {Badenes}, {Gaensler},
  {Gelfand}, {Hughes}, \& {Slane}}]{rakowski06}
{Rakowski}, C.~E., {et~al.} 2006, \apj, 646, 982

\bibitem[{{Reed} {et~al.}(1995){Reed}, {Hester}, {Fabian}, \&
  {Winkler}}]{reed95}
{Reed}, J.~E., {et~al.} 1995, \apj, 440, 706

\bibitem[{{Reich} {et~al.}(2003){Reich}, {Zhang}, \& {F{\"u}rst}}]{reich03}
{Reich}, W., {et~al.} 2003, \aap, 408, 961

\bibitem[{{Rho} {et~al.}(2001){Rho}, {Jarrett}, {Cutri}, \& {Reach}}]{rho01}
{Rho}, J., {et~al.} 2001, \apj, 547, 885

\bibitem[{{Rho} \& {Petre}(1998)}]{mixed98}
{Rho}, J. \& {Petre}, R. 1998, \apjl, 503, L167

\bibitem[{{Shelton} {et~al.}(1999){Shelton}, {Cox}, {Maciejewski}, {Smith},
  {Plewa}, {Pawl}, \& {R{\'o}{\.z}yczka}}]{shelton99}
{Shelton}, R.~L., {et~al.} 1999, \apj, 524, 192

\bibitem[{{Shelton} {et~al.}(2004){Shelton}, {Kuntz}, \& {Petre}}]{shelton04a}
{Shelton}, R.~L., {et~al.} 2004, \apj, 611, 906

\bibitem[{{Smith} {et~al.}(2001){Smith}, {Brickhouse}, {Liedahl}, \&
  {Raymond}}]{aped}
{Smith}, R.~K., {et~al.} 2001, \apjl, 556, L91

\bibitem[{{Snell} {et~al.}(2005){Snell}, {Hollenbach}, {Howe}, {Neufeld},
  {Kaufman}, {Melnick}, {Bergin}, \& {Wang}}]{snell05}
{Snell}, R.~L., {et~al.} 2005, \apj, 620, 758

\bibitem[{{Sturner} \& {Dermer}(1995)}]{sturner95}
{Sturner}, S.~J. \& {Dermer}, C.~D. 1995, \aap, 293, L17

\bibitem[{{Thielemann} {et~al.}(1996){Thielemann}, {Nomoto}, \&
  {Hashimoto}}]{thielemann96}
{Thielemann}, F.-K., {et~al.} 1996, \apj, 460, 408

\bibitem[{{Troja} {et~al.}(2006){Troja}, {Bocchino}, \& {Reale}}]{troja06}
{Troja}, E., {et~al.} 2006, \apj, 649, 258

\bibitem[{{Truelove} \& {McKee}(1999)}]{tm99}
{Truelove}, J.~K. \& {McKee}, C.~F. 1999, \apjs, 120, 299

\bibitem[{{van der Swaluw} {et~al.}(2004){van der Swaluw}, {Downes}, \&
  {Keegan}}]{swaluw04}
{van der Swaluw}, E., {et~al.} 2004, \aap, 420, 937

\bibitem[{{van Dishoeck} {et~al.}(1993){van Dishoeck}, {Jansen}, \&
  {Phillips}}]{vandish93}
{van Dishoeck}, E.~F., {et~al.} 1993, \aap, 279, 541

\bibitem[{{Wang} \& {Chevalier}(2002)}]{wang02}
{Wang}, C.-Y. \& {Chevalier}, R.~A. 2002, \apj, 574, 155

\bibitem[{{Wang} {et~al.}(1992){Wang}, {Asaoka}, {Hayakawa}, \&
  {Koyama}}]{wang92}
{Wang}, Z.~R., {et~al.} 1992, \pasj, 44, 303

\bibitem[{{Willingale} {et~al.}(2002){Willingale}, {Bleeker}, {van der Heyden},
  {Kaastra}, \& {Vink}}]{willingale02}
{Willingale}, R., {et~al.} 2002, \aap, 381, 1039

\bibitem[{{Woosley} \& {Weaver}(1995)}]{ww95}
{Woosley}, S.~E. \& {Weaver}, T.~A. 1995, \apjs, 101, 181

\bibitem[{{Zavlin}(2007{\natexlab{a}})}]{zavlin07b}
{Zavlin}, V.~E. 2007{\natexlab{a}}, \apjl, 665, L143

\bibitem[{{Zavlin}(2007{\natexlab{b}})}]{zavlin07a}
{Zavlin}, V.~E. 2007{\natexlab{b}}, \apss, 308, 297

\end{thebibliography}
